# DETERMINATION OF CIRCULAR POLARIZATION OF γ-QUANTA WITH energy > 10 MeV USING COMPTON POLARIMETER.

A.S. Arychev[*], A. P. Potylitsyn
*Tomsk Polytechnic University, pr. Lenina 30, 634034 Tomsk, Russian Federation*
M. N. Strikhanov
*Moscow Engineering Physical Institute, Kashirskoe shosse 31, 115409 Moscow, Russian Federation*

**Abstract**

On the KEK-ATF accelerator the proof-of-principle experiment will be carried out, in which is planned to obtain a circularly polarized beam of γ-quanta with energy < 60 MeV for the subsequent generation of longitudinal - polarized positrons. [1]

The circularly polarized γ-quanta with energy > 10 MeV are formed during Compton backscattering of circularly polarized laser photons on an electron beam with energy 1.26 GeV. The contribution of nonlinear processes for planned parameters of a laser flash [2] can be rather significant, that leads to deterioration of polarization of a resulting γ- beam. The direct method of measurement of γ-beam circular polarization will allow not only to estimate a correctness of the theoretical calculations, but also to use the obtained information for simulation of polarization parameters of positrons.

We offer to use a Compton polarimeter for this purpose where the recoil electrons after an interaction of a circularly polarized γ-quanta with a thin iron magnetized target are detected. The magnetic analysis of the recoil electrons allows to reject effectively the contribution of background processes. The geometry of a polarimeter was selected from a maximum of figure of merit $R^2 \cdot d\sigma/d\Omega$, where *R* is asymmetry of a Compton scattering process and $d\sigma/d\Omega$ is a cross section of the process. The obtained estimation of the scattered electron yield $\sim 10^{-7} e^-/photon$ for the 20 μm target thickness allows to estimate a necessary statistics for receiving of 10% error in circular polarization of quanta with energy ~15 MeV (~$10^4$ photon bunches with intensity relevant to second stage of the experiment [1]).

---

[*] Author to whom correspondence should be addressed. E-mail: s066002@chair12.phtd.tpu.edu.ru
[1] T. Hirose, K. Dobaski and Y. Kurihara, Nucl. Instr. and Meth. 455, p. 15 (2000).
[2] T. Okugi, Y. Kurihara and M. Chiba, Jpn. J. Appl. Phys. 35, p. 3677 (1996).

**Introduction**

A proof-of-principle experiment will be carried out on the KEK-ATF accelerator, where it is planned to obtain a circularly polarized beam of γ-quanta with the energy < 60 MeV for the subsequent generation of longitudinally - polarized positrons. [1]

Circularly polarized γ-quanta with the energy > 10 MeV are formed by Compton backscattering of circularly polarized laser photons in a 1.26 GeV electron beam. The contribution from nonlinear processes at the given laser flash parameters [2] can be rather appreciable, which leads to deterioration of the resulting γ- beam polarization.

The experimental measurements of the degree of circular polarization of a γ-beam will primarily allow one to verify the theoretical calculations (or the simulation results). Secondly, the experimental error of the direct method of measuring γ- beam circular polarization is likely to be less than the theoretical uncertainty. It should be mentioned that simulation of the $e^+ - e^-$ pair production process in an amorphous converter (taking into account the polarization states of the particles participating in the reaction) may be done to a much higher precision than that of nonlinear processes in a Gaussian beam of laser photons. Therefore, simulation with the known spectrum and polarization of a γ- beam will provide reliable information on polarization of a positron beam.

We offer to use a Compton polarimeter for the analysis of circular polarization of a γ-quantum with the energy $E_\gamma$ >10MeV. In an earlier experiment [3], a similar polarimeter was used to measure circular polarization of a γ-quantum with the energy E = 2.6 MeV. However, as the γ-quantum energy grows the efficiency of such a polarimeter is decreased. Furthermore, the time history of the γ- beam reported in the experiment [1] is not helpful in detecting the scattered γ-quanta with a sufficient rejection of the background processes. The proposed set-up of a Compton polarimeter is shown in Fig. 1. A circularly polarized γ-quantum beam passes through a thin ferromagnetic foil placed in a constant magnetic field. About 7 % electrons in the target are oriented by the spin along the field, whose reversal allows one to change the electron polarization sign.

**Compton polarimeter for the analysis of circular polarization of a γ- beam**

The Compton scattering process (two-particle reaction) is entirely determined by the angle of one of the outgoing particles (electron or photon). In the polarimeter scheme shown, the recoil electrons are to be detected by a magnetic analyzer. The analyzer field is selected such that the electron momentum corresponds to the outgoing angle $\theta_e$, at which the analyzer is located (Fig. 1). In this case, the contribution from background electrons from pair production in the nucleus field (three-particle reaction) will be largely suppressed. This polarimeter scheme allows one to detect Compton electrons during the spill of a few picosecond duration (time of one photon bunch passage through the target). In this case the response of the detector D (a Cherenkov counter) will correspond to the sum of responses of each photon.

---
[3] T. Sasahara "Positron Production from a Tungsten Single Crystal at the KEK 8-GeV Electron Linac", Graduate School of Science, Department of Physics, Tokyo Metropolitan University.

The cross section of the circularly polarized γ-quanta scattered on a longitudinaly - polarized electron has a well-known form

$$\frac{d\sigma}{d\omega} = \frac{1}{2} \cdot r_0^2 \cdot \frac{k^2}{k_0^2} \left\{ \left[ \frac{k_0}{k} + \frac{k}{k_0} - \sin^2\theta \right] + P_c \cdot P_e \cdot \left[ \left( \frac{k_0}{k} - \frac{k}{k_0} \right) \cdot \cos\theta \right] \right\} = \frac{d\sigma_0}{d\Omega} \cdot \{1 + P_c \cdot P_e \cdot R\}. \quad (1)$$

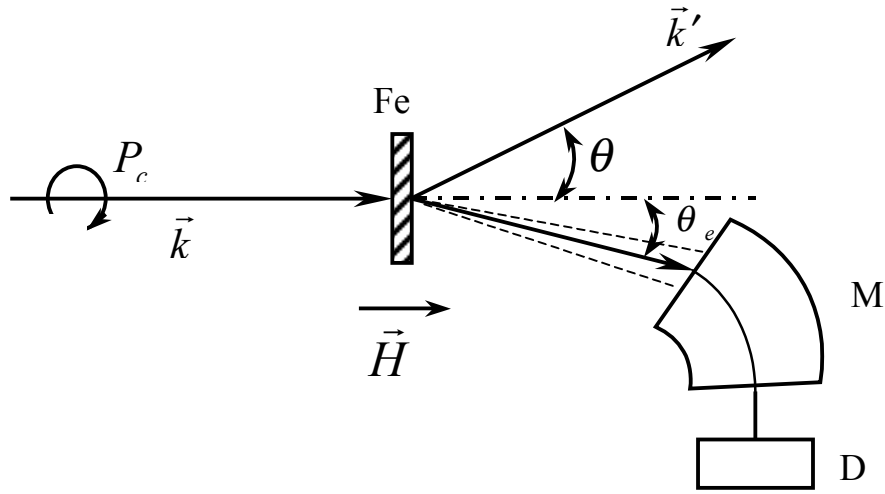

Fig. 1

In Eq. (1), $k_0(k)$ is the energy of the initial (scattered) quantum, $\theta$ is the photon scattering angle, $r_0 = \frac{e^2}{mc^2}$ is the classical electron radius, $P_c$ is the degree of circular polarization (for the right-hand polarization it is taken with the plus, for left-hand — with the minus), $P_e$ is the degree of longitudinal polarization of an electron (along the initial photon momentum), and $R$ is the asymmetry ratio.

Parameter $\frac{d\sigma_0}{d\Omega}$ in formula (1) denotes the Compton scattering cross-section of an unpolarized photon on an unpolarized electron

$$\frac{d\sigma_0}{d\Omega} = r_0^2 \cdot \left( \frac{k}{k_0} \right) \cdot \left\{ \frac{k_0}{k} + \frac{k}{k_0} - \sin^2\theta \right\}. \quad (2)$$

The relation between $k$ and $k_0$ is determined by the conservation law

$$k = \frac{k_0}{1+\frac{k_0}{mc^2}(1-\cos\theta)}.$$

The asymmetry ratio of the process is expressed as follows:

$$R = \frac{\left(\frac{k_0}{k}-\frac{k}{k_0}\right)\cdot\cos\theta}{\frac{k_0}{k}+\frac{k}{k_0}-\sin^2\theta}. \qquad (3)$$

In the proposed polarimeter, the quantity $P_e \cdot R$ determines the analyzing power of the device.

Figure 2 shows the differential cross section and asymmetry dependences on the scattered γ-quantum angle for the initial photon energy $k_0 = 30\cdot mc^2 \approx 15\,\text{MeV}$.

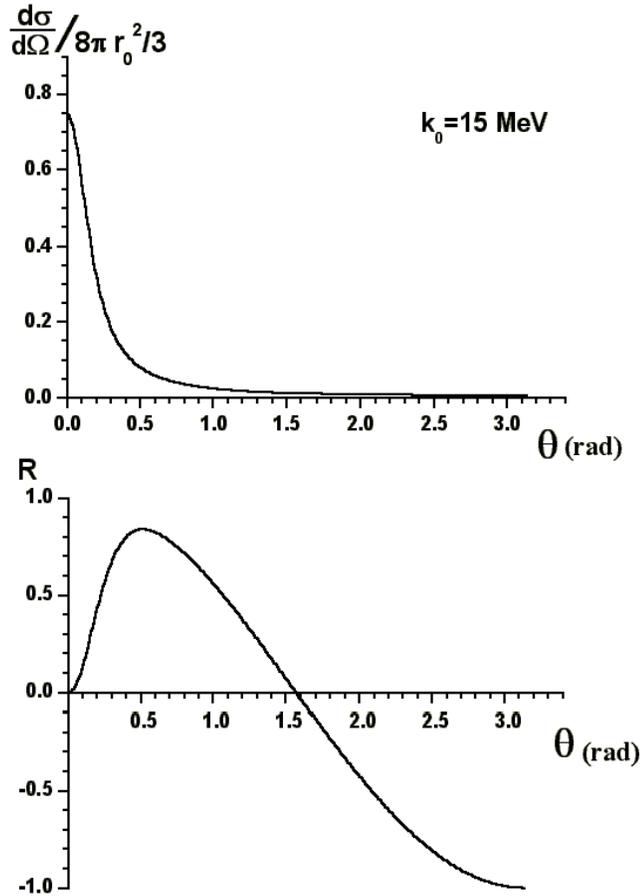

Fig. 2

Let us express equation (1) in terms of the recoil electron angle $\theta_e$ (see fig. 1), which is related to the photon angle $\theta$ as follows:

$$\cos^2\theta_e = (1+\omega)^2 \cdot \frac{1-\cos\theta}{2+\omega\cdot(\omega+2)\cdot(1-\cos\theta)}. \qquad (4)$$

Angle $\theta_e$ varies from $\frac{\pi}{2}$ to zero as $\theta$ goes from 0 to $\pi$ (fig. 3). In (4), use was made of the relation $\omega = k_0/mc^2$.

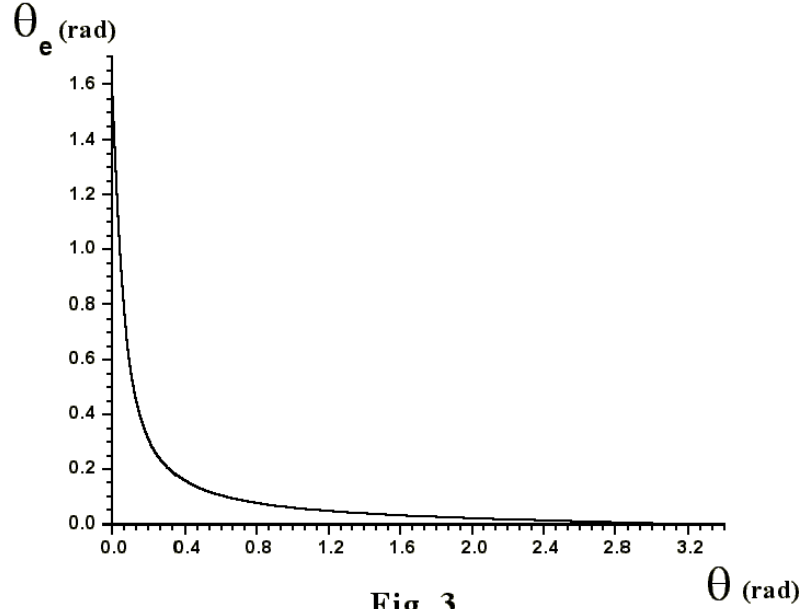

**Fig. 3**

Using (4), one can readily obtain an expression for the differential cross section and asymmetry in terms of the scattered electron angle

$$\frac{d\sigma}{d\Omega_e} = 4\pi \cdot r_0^2 \cdot \frac{(1+\omega)^2 \cdot \cos\theta_e}{(1+2\omega+\omega^2\sin^2\theta_e)^2} \times$$
$$\times\left[1 + \frac{2\omega^2 \cdot \cos^4\theta_e}{(1+2\omega+\omega^2\sin^2\theta_e)\cdot(1+\omega(\omega+2)\cdot\sin^2\theta_e)} - \frac{2\cdot(1+\omega)^2 \cdot \sin^2\theta_e \cos^2\theta_e}{(1+\omega(\omega+2)\cdot\sin^2\theta_e)^2}\right]; \qquad (5)$$

$$R = \frac{\left(\dfrac{k_0}{k} - \dfrac{k}{k_0}\right)\cdot\left[1 - \dfrac{2\cdot\cos^2\theta_e}{(1+\omega)^2 - \cos^2\theta_e \cdot(\omega^2+2\cdot\omega)}\right]}{\dfrac{k_0}{k} + \dfrac{k}{k_0} - \dfrac{4\cdot\cos^2\theta_e \cdot\left[(1+\omega)^2 - \cos^2\theta_e \cdot(\omega^2+2\cdot\omega)\right] + 4\cdot\cos^4\theta_e}{(1+\omega)^4 - 2\cdot(1+\omega)^2 \cdot \cos^2\theta_e \cdot(\omega^2+2\cdot\omega) + \cos^4\theta_e \cdot(\omega^2+2\cdot\omega)^2}}$$

where $k = \dfrac{k_0}{1 + \omega \cdot \left( \dfrac{2 \cdot \cos^2 \theta_e}{(1+\omega)^2 - \cos^2 \theta_e \cdot (\omega^2 + 2\omega)} \right)}$.

Figure 4 shows the differential cross section and asymmetry dependences on the scattering angle $\theta_e$ for $\omega = 30$.

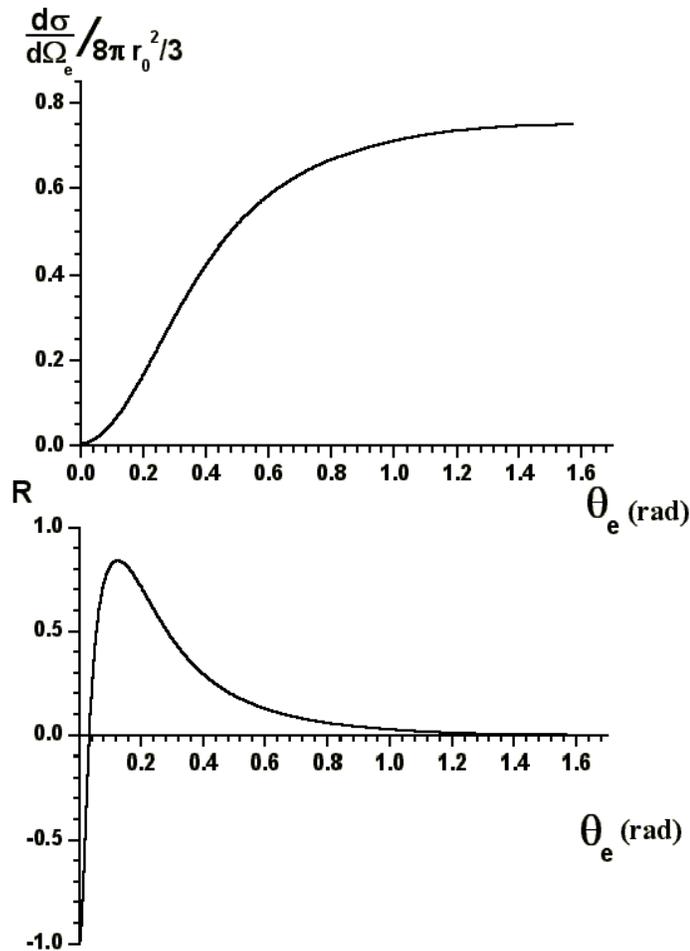

Fig. 4

The statistical error of polarization measurements is determined by so-called figure of merit $\dfrac{d\sigma}{d\Omega} \cdot R^2$. Let us choose the kinematics of the Compton scattering process corresponding to the maximum figure of merit (fig. 5).

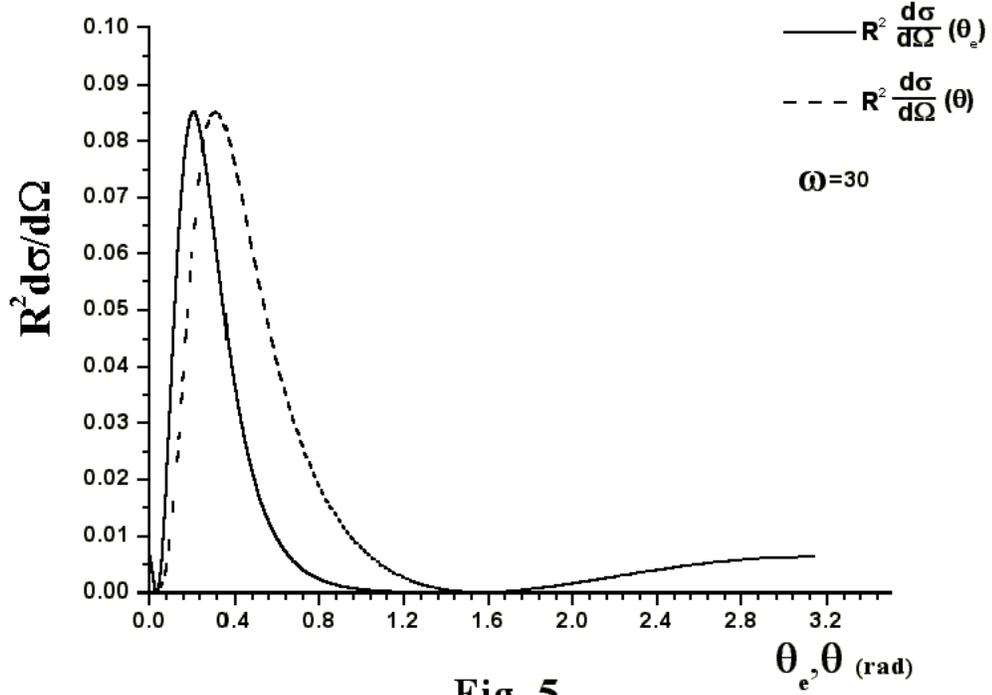

**Fig. 5**

For the photon energy $\omega = 30$ (15 MeV), the angle $\theta_e$ corresponding to the peak in the curve is equal to 0.208 radian. For this geometry, the differential cross section $\frac{d\sigma}{d\Omega_e}$ is as large as $1.18 \cdot 10^{-25}\, cm^2$ and the recoil electron energy is $\varepsilon = 17.34 \cdot mc^2 = 8.86$ MeV.

The number of recoil electrons $N_e$ per an initial photon is determined by a well-known equation

$$N = \tfrac{d\sigma}{d\Omega} \cdot \Delta\Omega \cdot n \cdot t, \qquad (6)$$

where $\Delta\Omega$ is the solid angle, which is measured by the analyzer aperture, n is the concentration of electrons in the target, and t is the target thickness.

For the target thickness $t = 20\,\mu m$, the yield of recoil electrons for an analyzer with $\Delta\Omega = 4 \cdot 10^{-4}$ is equal to $N_e = 7.8 \cdot 10^{-8}\ e^-/photon$.

It should be emphasized that at this quantum energy the contribution from the electrons formed through pair production is quite significant. The cross section of this process in the nucleus field charge Z is written as follows:

$$\frac{d\sigma}{d\Omega} = Z^2 \cdot \alpha \cdot \frac{r_0^2}{k_0^3 \cdot 2\pi} \left\{ (\varepsilon_1^2 + \varepsilon_2^2)\cdot(3+2\Gamma) + 2\varepsilon_1^2\varepsilon_2^2 \cdot \left(1 + 4\frac{\gamma^2\theta_e^2}{(1+\gamma^2\theta_e^2)^2} \cdot \left( \ln\left[\frac{111\cdot(1+\gamma^2\theta_e^2)}{\sqrt[3]{Z}}\right] - 2 \right) \right) \right\} \times$$

$$\times \frac{\gamma^2}{(1+\gamma^2\theta_e^2)^2} \cdot \frac{\Delta\varepsilon_1}{\varepsilon_1}, \tag{7}$$

where $\Gamma = \ln\left[\frac{111 \cdot Z^{-\frac{1}{3}}}{\xi}\right] - 2$, $\xi = \frac{1}{1+(\gamma \cdot \theta_e)^2}$,

$\varepsilon_1$, $\varepsilon_2$ - are the electron and positron energies, which in the case in question are equal to $\varepsilon_1 = 8.86$ MeV, $\varepsilon_2 = 5.45$ MeV.

The number of background electrons per an initial quantum with the energy $\omega = 30$ in the same target is equal to $N = 5.6 \cdot 10^{-9}$, which is by an order of magnitude less in comparison with the Compton scattering electrons.

Polarization experiments generally measure the assymetry

$$\Sigma = \frac{y_+ - y_-}{y_+ + y_-} \tag{8}$$

Here $y_{+,-}$ is the reaction yield for the opposite orientations of the electron spin in the target. For Compton scattering, we have

$$y_+ = \frac{1}{2} \cdot N_e \cdot N_{ph} \cdot (1 + P_c \cdot \overline{R})$$

$$y_- = \frac{1}{2} \cdot N_e \cdot N_{ph} \cdot (1 - P_c \cdot \overline{R})$$

Here $N_{ph}$ is the total number of quanta per exposure, during which asymmetry $\Sigma$ is measured and $\overline{R} \approx 0.07 \cdot R$ is the analyzing power of Compton scattering in a magnetized foil.

The degree of the photon circular polarization is determined from the experimentally measured value (8)

$$P_c = \frac{\Sigma}{\overline{R}}.$$

It can be shown that at $P_c \to 1$, the statistical error of the measured polarization is determined by the formula

$$\frac{\Delta P}{P} \sim \frac{1}{\sqrt{N_e \cdot \overline{R}^2}}$$

The quantity $N_e \cdot \overline{R}^2$ (so-called normalized statistics) significantly exceeds the statistics necessary to obtain the same error while measuring unpolarized beams and targets. In this case $R \approx 0.7$, hence $\overline{R} = P_e \cdot R \approx 0.05$. For the error as low as 10%, $\Delta P_e / P_e \sim 0.1$, it is necessary to detect $\sim 4 \cdot 10^4$ electrons, which requires the exposure $N_{ph} \sim 10^{12}$. A similar exposure for the second stage of the experiment /1/ may be achieved for $\sim 10^5$ photon bunches passing through the target.